\documentclass[floats,floatfix,showpacs,amssymb,prd,superscriptaddress,twocolumn,aps]{revtex4}
\usepackage{graphicx, epsfig, amssymb} %include figure files
\usepackage{amsmath, amsfonts}
 \usepackage{gensymb}
\usepackage{bm} %include bold math: \bm{} creates bold letters in math mode

\usepackage[linktocpage]{hyperref}
\usepackage[caption=false]{subfig}
\usepackage[usenames]{color}

\def\be{\begin{equation}}
\def\ee{\end{equation}}
\def\beq{\begin{eqnarray}}
\def\eeq{\end{eqnarray}}

\newcommand{\bea}{\begin{eqnarray}}
\newcommand{\eea}{\end{eqnarray}}
\newcommand{\ben}{\begin{enumerate}}
\newcommand{\een}{\end{enumerate}}
\newcommand{\bi}{\begin{itemize}}
\newcommand{\ei}{\end{itemize}}

\newcommand{\nn}{\nonumber}

\pdfoutput=1

\begin{document}

%%%%%%%%%%%%%%%%%
%%%   TITLE   %%%
%%%%%%%%%%%%%%%%%
\title{Kerr black holes with synchronised hair: \\  an analytic  model and dynamical formation}

 \author{Carlos~A.~R.~Herdeiro}
   \affiliation{
   Departamento de F\'\i sica da Universidade de Aveiro and CIDMA, 
   Campus de Santiago, 3810-183 Aveiro, Portugal.
 }

 \author{Eugen Radu}
   \affiliation{
   Departamento de F\'\i sica da Universidade de Aveiro and CIDMA, 
   Campus de Santiago, 3810-183 Aveiro, Portugal.
 }

%%%%%%%%%%%%%%%%
%%%   DATE   %%%
%%%%%%%%%%%%%%%%

\date{June 2017}

%%%%%%%%%%%%%%%%%%%%
%%%   ABSTRACT   %%%
%%%%%%%%%%%%%%%%%%%%
\begin{abstract}
East and Pretorius~\cite{East:2017ovw} have successfully evolved, using fully non-linear numerical simulations, the superradiant instability of the Kerr black hole (BH) triggered by a massive, complex vector field. Evolutions terminate in stationary states of a vector field condensate synchronised with a rotating BH horizon. We show these end points are fundamental states of Kerr BHs with synchronised Proca hair. Motivated by the ``experimental data" from these simulations we suggest a \textit{universal} ($i.e.$ field-spin independent), \textit{analytic} model for the subset of  BHs with sychronised hair that possess a \textit{quasi}-Kerr horizon, applicable in the weak hair regime. Comparing this model with fully non-linear numerical solutions of BHs with synchronised scalar or Proca hair, we show the model is accurate for hairy BHs that may emerge dynamically from superradiance, whose domain we identify.
\end{abstract}

%%%%%%%%%%%%%%%%
%%%   PACS   %%%
%%%%%%%%%%%%%%%%


%%%%%%%%%%%%%%%%%%%%%%
%%%   MAKE TITLE   %%%
%%%%%%%%%%%%%%%%%%%%%%

\maketitle
%%%%%%%%%%%%%%%%%%%%%%%%%%%%%%%%%%%%%%%%%%%%%%%%%%%%%%%%%%%%%%%%%%%%%
%%%%%%%%%%%%%%%%%%%%%%%%%%%%%%%%%%%%%%%%%%%%%%%%%%%%%%%%%%%%%%%%%%%%%
\noindent{\bf {\em Introduction.}} A 50 year-old lingering question in black hole (BH) physics has been the endpoint of the Kerr BH superradiant instability~\cite{Press:1972zz}, triggered by massive, bosonic fields~\cite{Damour:1976kh,Zouros:1979iw,Detweiler:1980uk,Brito:2015oca}.  Apart from its theoretical interest, this instability offers an unexpected opportunity for testing the existence of ultra-light bosonic particles suggested by beyond the standard model scenarios, $e.g.$~\cite{Arvanitaki:2009fg}.  BHs effectively become \textit{particle detectors}~\cite{Arvanitaki:2010sy}, creating a remarkable synergy between strong gravity, particle physics and astrophysics, testable by ongoing/future gravitational waves and electromagnetic observations (see $e.g$~\cite{Arvanitaki:2010sy,Pani:2012vp,Witek:2012tr,Yoshino:2012kn,Yoshino:2013ofa,Pani:2013hpa,Arvanitaki:2014wva,Brito:2014wla,Cunha:2015yba,Yoshino:2015nsa,Vincent:2016sjq,Arvanitaki:2016qwi,Ni:2016rhz,Baryakhtar:2017ngi,Zhou:2017glv,Franchini:2016yvq,Fujita:2016yav}).

In a recent breakthrough, East and Pretorius~\cite{East:2017ovw} reported long term numerical evolutions of this instability, using a Proca field to trigger it. Their evolutions lead to equilibrium states wherein the BH horizon angular velocity, $\Omega_H$, \textit{synchronises} with the phase angular velocity of the (complex) Proca field. This suggests these endpoints belong to the family of \textit{Kerr BHs with synchronised Proca hair} (KBHsPH), previously constructed as fully non-linear, stationary solutions of the Einstein-Proca system~\cite{Herdeiro:2016tmi}.

In the present letter we establish the equilibrium states obtained dynamically in~\cite{East:2017ovw} correspond to 
\textit{fundamental states} of KBHsPH, making these BHs the first dynamical counter example to the no-hair conjecture~\cite{Ruffini:1971}, in general relativity, with a simple, physically reasonable matter content. Moreover, we identify which of these hairy BHs can be endpoints of superradiance.

More generically, for a subset of BHs with synchronised bosonic hair, including all solutions that may dynamically form from the superradiant instability, we propose an analytic model based on the hypothesis that the horizon is \textit{quasi-}Kerr. This holds when the BH hair is weak, but also even for considerably hairy BHs when the matter field is dilute. Within this model, explicit, universal ($i.e$ field-spin independent) formulas for physical quantities are presented, in terms of a ``hairiness" parameter. These provide an analytic handle for further studies of these solutions in this interesting regime.

%%%%%%%%%%%%%%%%%%%%%%
%%%%%%%%%%%%%%%%%%%%%%
\noindent{\bf {\em The \textit{quasi}-Kerr horizon (qKH) model.}} 
%%%%%%%%%%%%%%%%%%%%%%
%%%%%%%%%%%%%%%%%%%%%%
Consider a stationary, axisymmetric, asymptotically flat, BH with synchronised hair of a $generic$ bosonic matter field $\psi$, with field mass $\mu$. Synchronisation means that $\Omega_H=w/m$ where $\psi\sim e^{-i(wt-m\phi)}\dots $, $\partial/\partial t$, $\partial/\partial \phi$ are the Killing vector fields associated to stationarity and axi-symmetry, respectively, $m\in \mathbb{Z}^+$ and $w$ is the field's frequency.
Known examples have been constructed for both scalar~\cite{Herdeiro:2014goa,Herdeiro:2015gia,Kleihaus:2015iea,Herdeiro:2015tia} and vector matter~\cite{Herdeiro:2016tmi}. The ADM mass and angular momentum are $M$ and $J$, while the corresponding horizon  data are
$M_H$, $J_H$ (computed as Komar integrals) together with the area, $A_H$, temperature, $T_H$  and $\Omega_H$.
The corresponding matter field data obey (see $e.g$~\cite{Herdeiro:2016tmi})
\begin{eqnarray}
\label{Komar}
M_{(\psi)}=M-M_{H},~~
J_{(\psi)}=J-J_{H} \ .
\end{eqnarray}
The following Smarr relation also holds
%\begin{eqnarray}
%\label{smarr}
$M=\frac{1}{2} T_H A_H +2\Omega_H [J-J_{(\psi)}]+ M_{(\psi)},$
%\end{eqnarray}
or, equivalently
\begin{eqnarray}
\label{rel-hor}
M_H=\frac{1}{2} T_H A_H+2 \Omega_H J_H.
\end{eqnarray}
Furthermore, the solutions satisfy the first law of thermodynamics
%\begin{eqnarray}
%\label{first-law}
$dM=\frac{1}{2} T_H dA_H +\Omega_H dJ.$
%\end{eqnarray}

It is convenient to define the aforementioned physical quantities normalised by the ADM mass,
\begin{equation}
\label{quantities1}
j\equiv  \frac{J}{M^2},~~
a_H \equiv \frac{A_H}{16\pi M^2},~~w_H\equiv \Omega_H M,~~t_H\equiv 8\pi M T_H \ ,
\end{equation}
and to introduce two measures of the 'hairiness', which provide, respectively, 
the fraction of energy (angular momentum) in the matter field
\begin{eqnarray}
\label{quantities2}
p\equiv \frac{M_{(\psi)}}{M} \ ,~~
q\equiv \frac{J_{(\psi)}}{J}\ .
\end{eqnarray}

In the absence of hair ($p=0=q$), $i.e$ for Kerr,
\begin{eqnarray}
\label{Kerr1}
j=\frac{4w_H}{1+4w_H^2},~~a_H=\frac{1}{1+4 w_H^2},~~t_H=1-4w_H^2,
\end{eqnarray}
with
%\begin{eqnarray}
%\label{Kerr11}
$0\leqslant w_H \leqslant {1}/{2}$,
%\end{eqnarray}
the limits corresponding the Schwarzschild solution and extremal Kerr, respectively.

In the presence of hair ($p\neq 0 \neq q$) we assume the BH horizon is \textit{quasi-}Kerr. We expect this to hold in \textit{the weak hair regime},  $p,q\ll 1$~\footnote{When $p,q\rightarrow 0$, the hair becomes a stationary cloud on the Kerr background. For the scalar case these clouds were first discussed in~\cite{Hod:2012px} -- see also~\cite{Hod:2013zza,Herdeiro:2014goa,Hod:2016yxg,Hod:2016lgi}.}.
Under this assumption, we introduce the \textit{quasi-Kerr horizon} (qKH) model approximating the  horizon quantities by those of a Kerr BH, \textit{but with the replacements} $(M,J)\rightarrow(M_H,J_H)$:
\begin{eqnarray}
\label{i1}
&&\Omega_H =\frac{M_H}{2J_H} \left(1-\chi \right), 
%\\
%\label{i2}
\\
\label{i3}
&&A_H=8\pi M_H^2 
\left(
1+\chi
\right), \ \ T_H=\frac{\chi}{4\pi M_H \left( 1+\chi \right)},
\end{eqnarray}
where $\chi\equiv \sqrt{1-{J_H^2}/{M_H^4}}$. For the Kerr family, these formulas are exact, but for hairy BHs they are just an approximation.

For hairy BHs \eqref{rel-hor} still holds;  from (\ref{Komar}) 
together with the first law of thermodynamics and the qKH model, we find
that the matter field mass and angular momentum obey
\begin{eqnarray}
\label{1st}
dM_{(\psi)}=\Omega_H dJ_{(\psi)}~.
\end{eqnarray}

We formally integrate (\ref{1st}) treating $\Omega_H$ as an input parameter. This is justified regarding the hairy BH as a composed system of a horizon plus a matter distribution. Then $\Omega_H$ characterises the horizon subsystem which is external (albeit coupled) to the ``hair" subsystem. Then
\begin{eqnarray}
\label{rel1}
M_{(\psi)}= \Omega_H  J_{(\psi)},~~ i.e. ~~M-M_{H}= \Omega_H (J-J_{H})~.
\end{eqnarray}
An equivalent form of (\ref{rel1})
reads
\begin{eqnarray}
\label{rel2}
p=w_H j q.
\end{eqnarray}

Relations  (\ref{i1}),  (\ref{rel2}) 
give two constraints for the four variables $(p,q,w_H,j)$.
We may choose $p,w_H$ as the independent (control) parameters yielding the simple expressions
\begin{equation}
\label{expr1}
%&&
q= p\frac{1+4(1-p)^2w_H^2}{p+4(1-p)^2w_H^2}, \ \
% \\
 %\label{expr2}
%&&
 j=\frac{p+4(1-p)^2w_H^2}{w_H(1+4(1-p)^2w_H^2)},
\end{equation}
which are, thus, predictions of the qKH model that shall be compared against the fully non-linear (numerical) solutions in the next section. Similarly, 
\begin{eqnarray}
\label{expr3}
%&&
a_H= \frac{ (1-p)^2 }{1+4(1-p)^2w_H^2}, \ \ \ 
 %\\
 %\label{expr4}
%&&
t_H=\frac{1-4(1-p)^2w_H^2}{1-p}.
\end{eqnarray}
Observe the Kerr relations \eqref{Kerr1} are recovered as $p=0$.

\begin{figure}[t!]
\begin{center}
\includegraphics[width=0.45\textwidth]{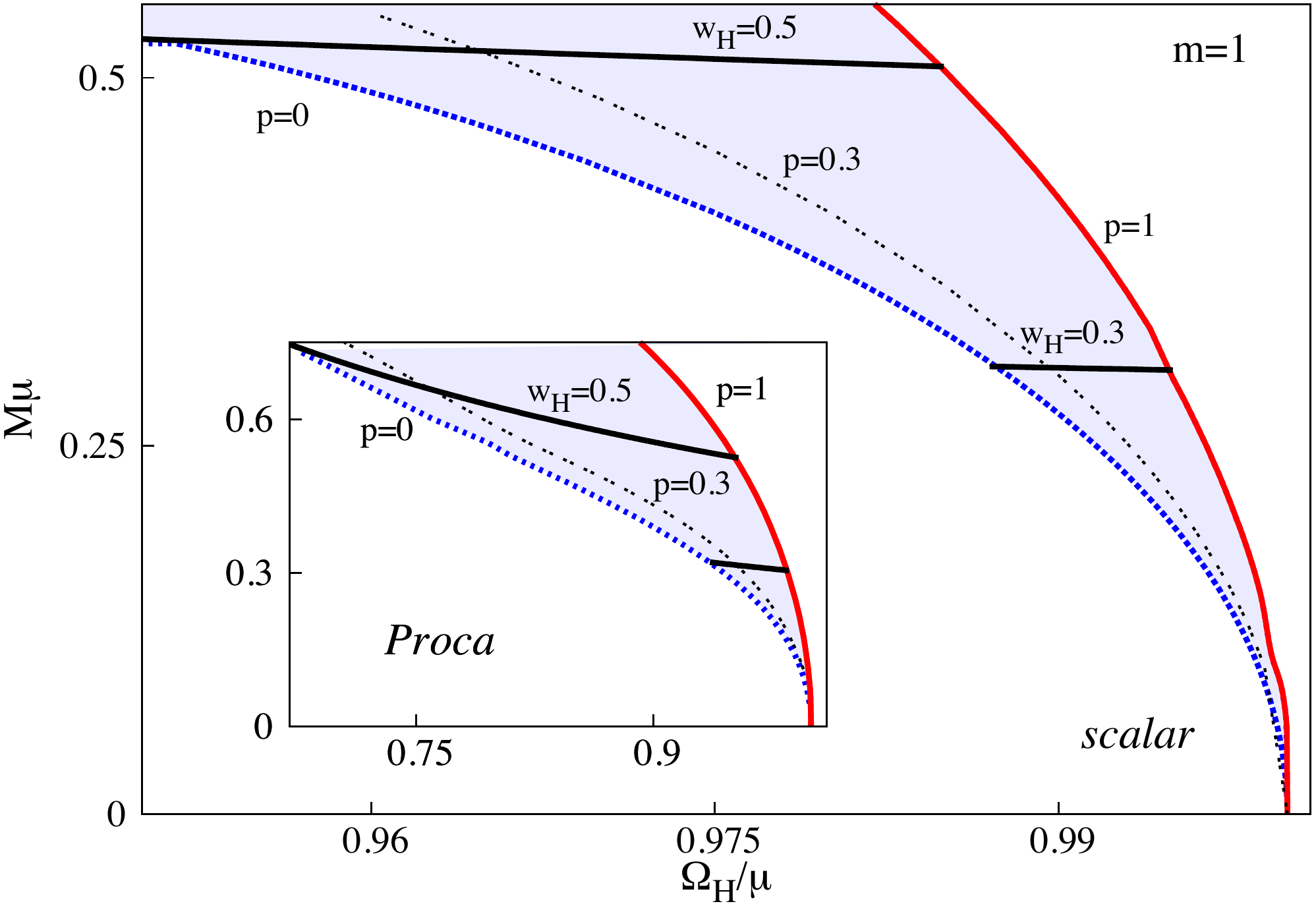}
\end{center}
\caption{\small Part of the domain of existence of Kerr BHs with scalar (main panel) or Proca (inset) hair (shaded blue region) in a $M$ $vs.$ $\Omega_H$ diagram, both in units of $\mu$. The Kerr limit -- existence line ($p=0$, blue dotted line) --   and the solitonic limit $(p=1$, red solid line) are shown, together with the line $p=0.3$ and the lines $\omega_H=0.3;0.5$, in both panels.}
\label{domain}
\end{figure}

%%%%%%%%%%%%%%%%%%%%%%
%%%%%%%%%%%%%%%%%%%%%%
\noindent{\bf {\em qKH model $vs.$ numerical solutions.}} 
%%%%%%%%%%%%%%%%%%%%%%
%%%%%%%%%%%%%%%%%%%%%%
To test the qKH model we compare it with fully non-linear (numerical) BHs with synchronised hair, which are solutions of the corresponding Einstein-matter system. To test universality, we shall consider both the scalar solutions constructed in~\cite{Herdeiro:2014goa,Herdeiro:2015gia} and Proca solutions. The latter belong to the family introduced in~\cite{Herdeiro:2016tmi} and are described in the Appendix. The corresponding part of the domain of existence to be explored is shown in Fig.~\ref{domain}~\footnote{The numerical results herein
are for $m=1$ being obtained from several thousands of
solution points for each type of hairy BHs, 
with typical error estimates $<10^{-3}$.}.

Fig.~\ref{domain} shows the neighbourhood of the \textit{existence line} wherein the hairy BHs reduce to vacuum Kerr ($p=0$). We shall be particularly interested in the hairy BHs that may arise from the growth of the superradiant instability of Kerr. Then, thermodynamics imposes an upper bound of $p\leqslant 1-1/\sqrt{2}\simeq 0.29$. The corresponding BHs exist in the \textit{allowed strip} between the $p=0$ and $p=0.3$ lines in~Fig.~\ref{domain}. The complete domain of existence can be found in~\cite{Herdeiro:2014goa} (Fig.~\ref{migration} below) for the scalar (Proca) case.

 In Fig.~\ref{colours} we test the qKH model by 
comparing the model's results for $\mathcal{Y}=\{j,a_H,t_H\}$
with the data from the numerical solutions.
It shows the relative errors 
$|1-\mathcal{Y}^{(th)}/\mathcal{Y}^{(num)}|$
for the allowed strip  with $p\in [0,0.3]$, 
$\omega_H\in [0,0.5]$.
This analysis shows that for the solutions with $j<1$, whose importance will be detailed below, the relative errors are below the percent level for all physical quantities, and typically lower for the scalar case. The errors are still only at a few percent level when $p\sim 0.3$ and for low values of $\omega_H$,  the relative errors are below 1\%, even for $p\sim 0.3$. We interpret this unexpected accuracy in the description as a consequence of a separation of scales. Indeed, $\omega_H<\mu M$, and  $M$ (the gravitational radius of the BH),  and $\mu$ (the inverse Compton wavelength of the massive bosonic field) are the two fundamental scales of the problem. When the dimensionless coupling $M \mu$ 
is small -- often called the \textit{Newtonian regime} -- the bosonic field is diluted, even though it may correspond to a large fraction of the total spacetime energy~\footnote{Such property, already observed in~\cite{Brito:2014wla}, is not found, as far as we know, in other models of hairy BH solutions.}. In this regime it is expectable that the BH horizon is not strongly affected by the (mostly far away) bosonic field distribution, being \textit{quasi-}Kerr. This type of separation of scales is often used to alleviate the complexity of physical problems. Two examples in the context of strong gravity appear in the study of spinning BH binaries~\cite{Gerosa:2015tea} and the blackfold approach to higher dimensional BHs~\cite{Emparan:2009cs}. 

\begin{widetext}

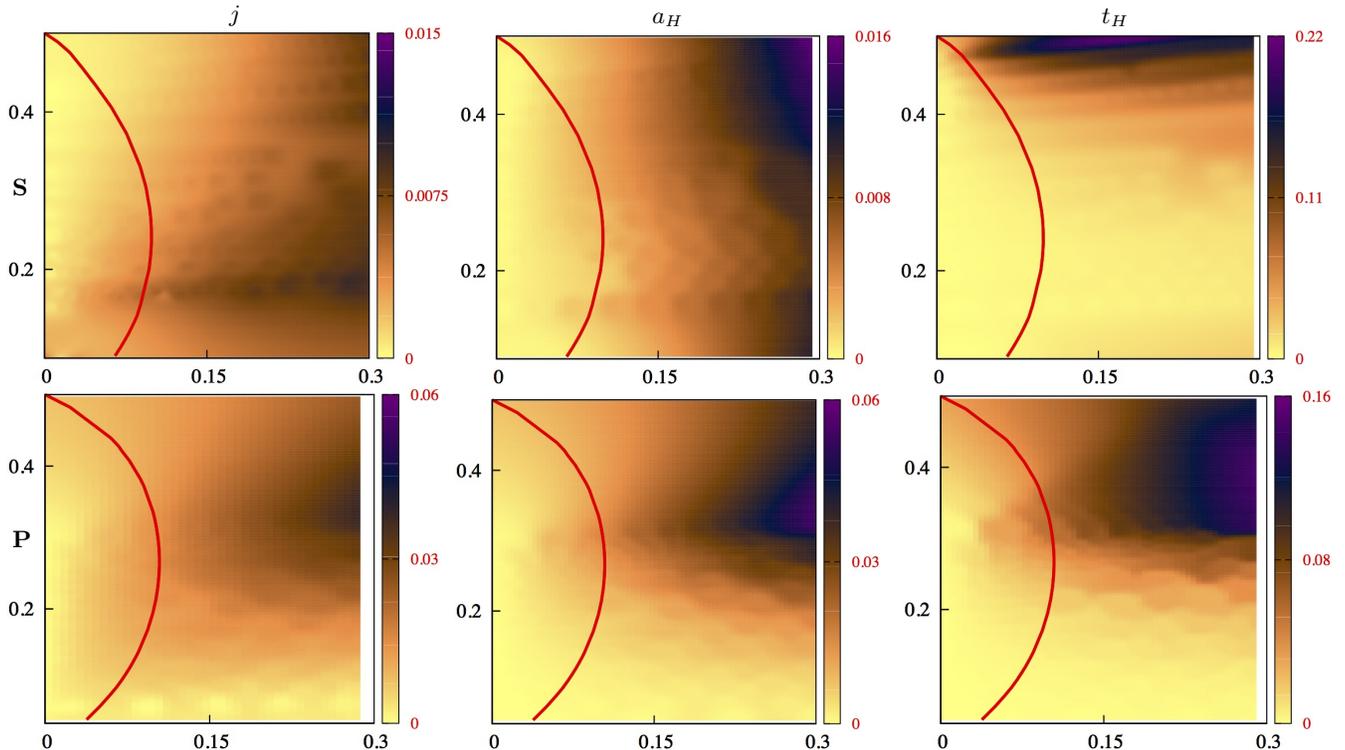
\begin{figure}[ht]
\begin{center}
 \includegraphics[width=0.33\textwidth]{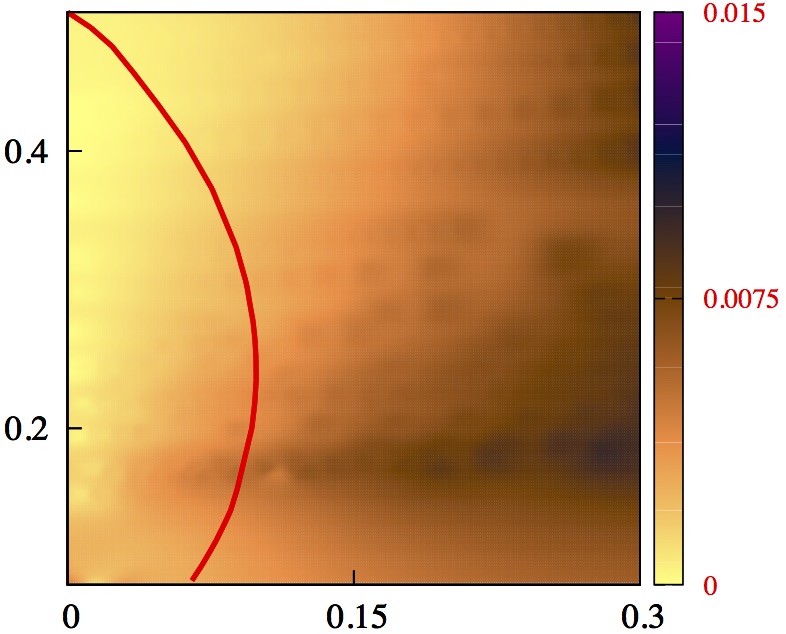}
\includegraphics[width=0.32\textwidth]{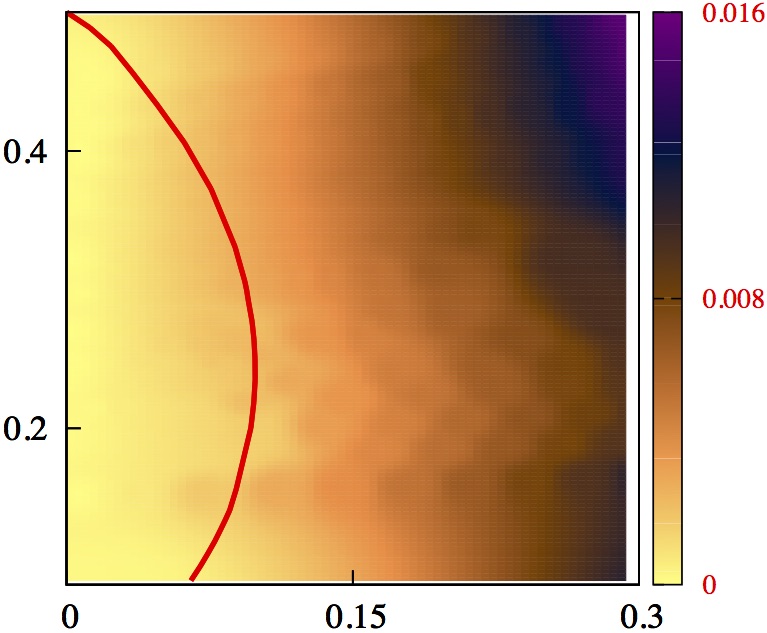}
\includegraphics[width=0.315\textwidth]{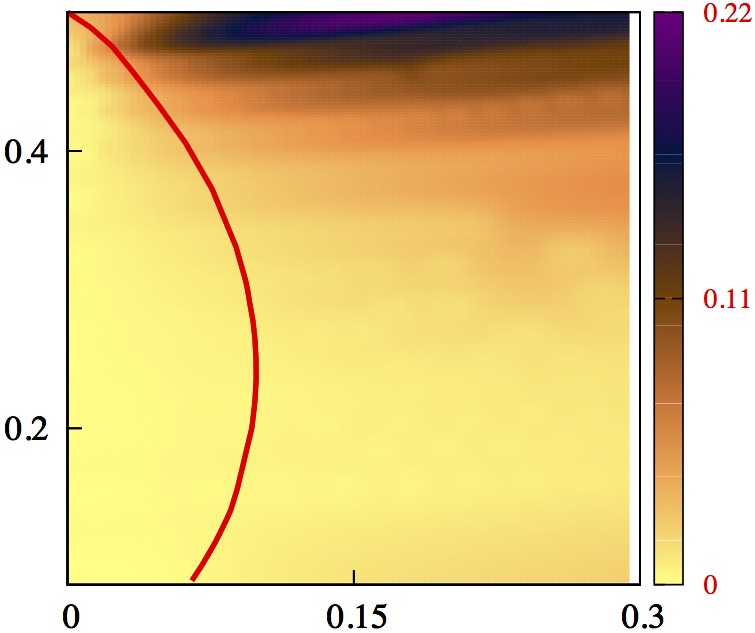} \\
\ \ \includegraphics[width=0.32\textwidth]{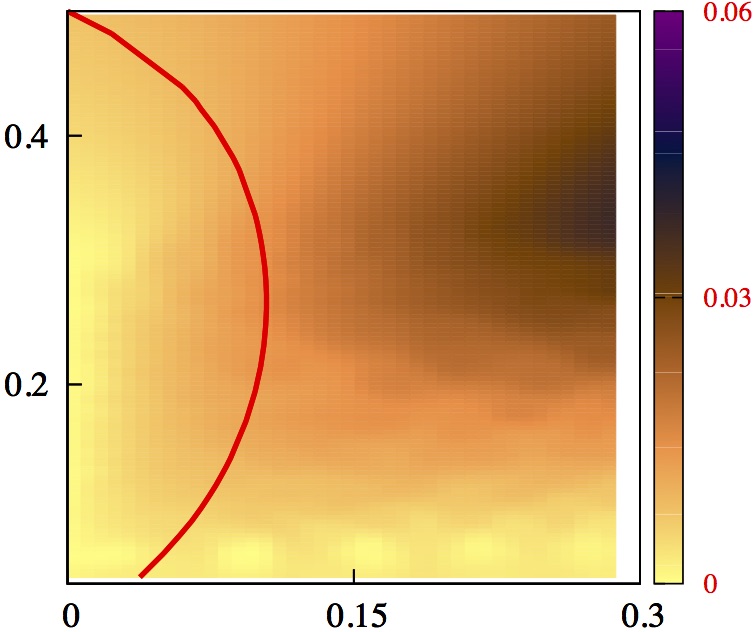} \ 
\includegraphics[width=0.315\textwidth]{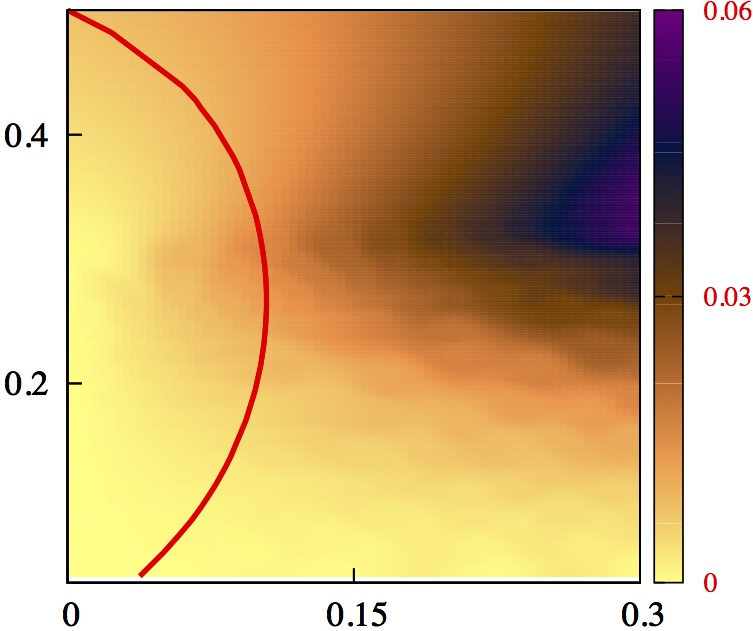} \ \
\includegraphics[width=0.318\textwidth]{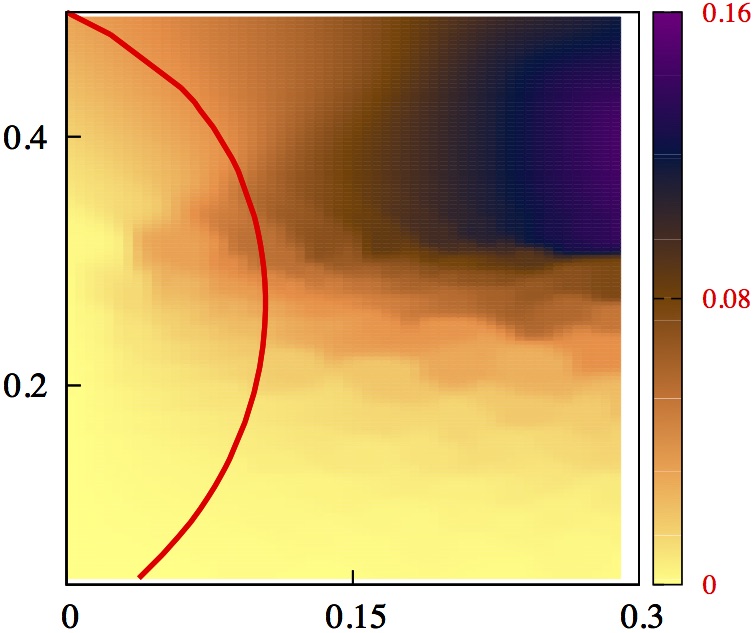}
\begin{picture}(0,0)
\put(-90,275){$t_H$}
\put(-260,275){$a_H$}
\put(-502,210){${\bf S}$}
\put(-502,77){${\bf P}$}
\put(-420,276){$ j$}
\end{picture}
\end{center}
\caption{\small The relative errors are shown in the allowed strip on the $(p,w_H)$-plane for the reduced angular momentum $j$ (left panels), area $a_H$ (middle panels) and temperature $t_H$ (right panels) for both BHs with scalar ({\bf S} - top panels) and Proca ({\bf P} - bottom panels) hair. The red lines are the set of solutions with $j=1$; solutions to the left of these lines have $j<1$.}
\label{colours}
\end{figure}

\end{widetext}

\begin{figure}[t!]
\begin{center}
\includegraphics[width=0.47\textwidth]{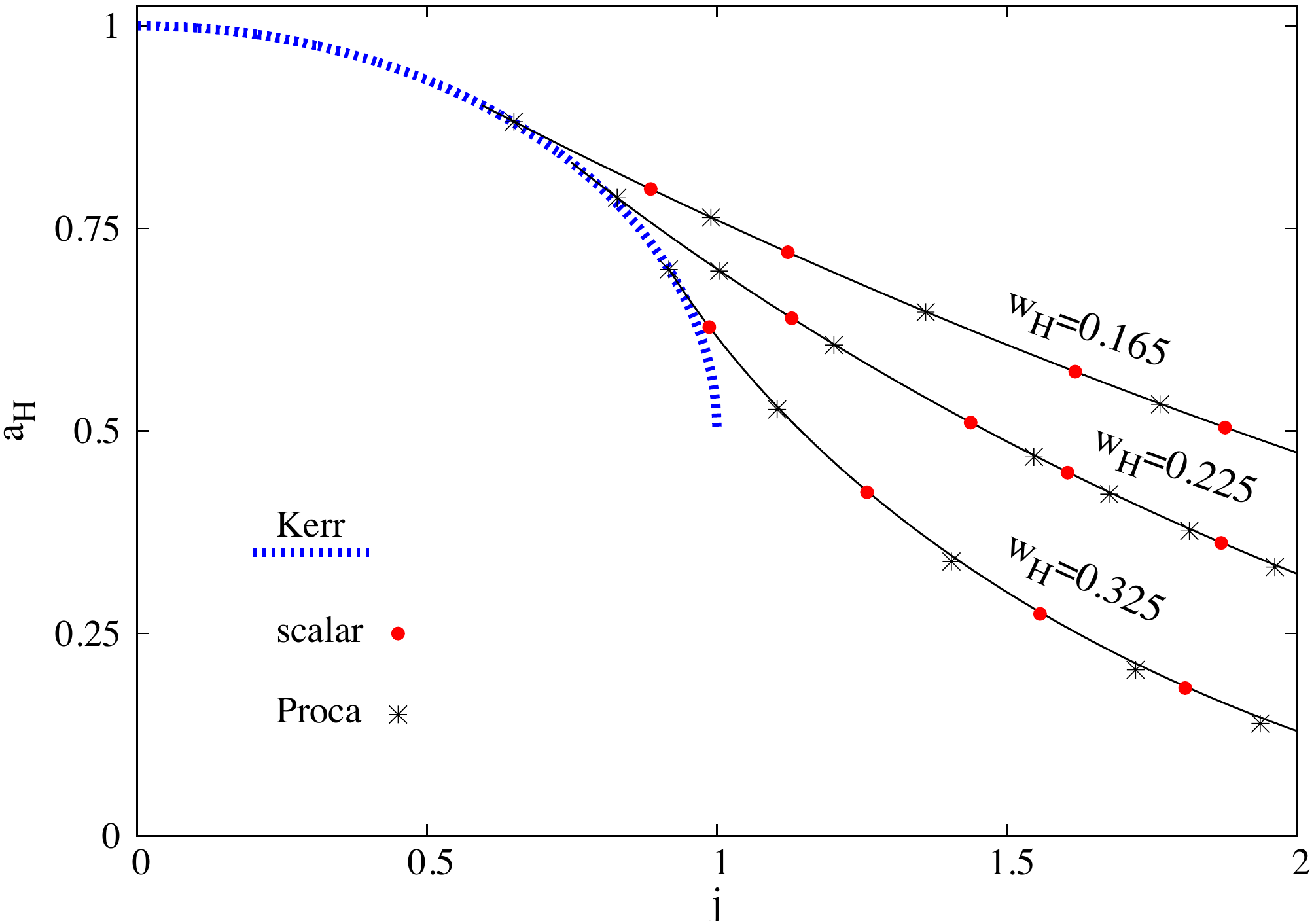}
\end{center}
\caption{\small $j$ $vs.$  $a_H$ diagram.
The black stars are Proca numerical solutions; the red points are for scalar hairy BHs.
The curves correspond to the analytical model for fixed $\omega_H$ values.}
\label{points}
\end{figure}

The analysis of Fig.~\ref{colours} validates the qKH model for the region where $j<1$. But we emphasise that even for $j>1$ the model can be accurate, especially for small $\omega_H$. This is illustrated in Fig.~\ref{points} where we exhibit  a $a_H$ $vs.$ $j$ plot. Even for $j$ approaching 2 (thus for non-Kerr BHs) the analytic model fits well the numerical points,  along lines of constant $\omega_H$, for \textit{both} the scalar and Proca case, supporting its (matter field spin) universality.

%\begin{widetext}

%\end{widetext}

\begin{figure}[ht]
\begin{center}
\includegraphics[width=0.48\textwidth]{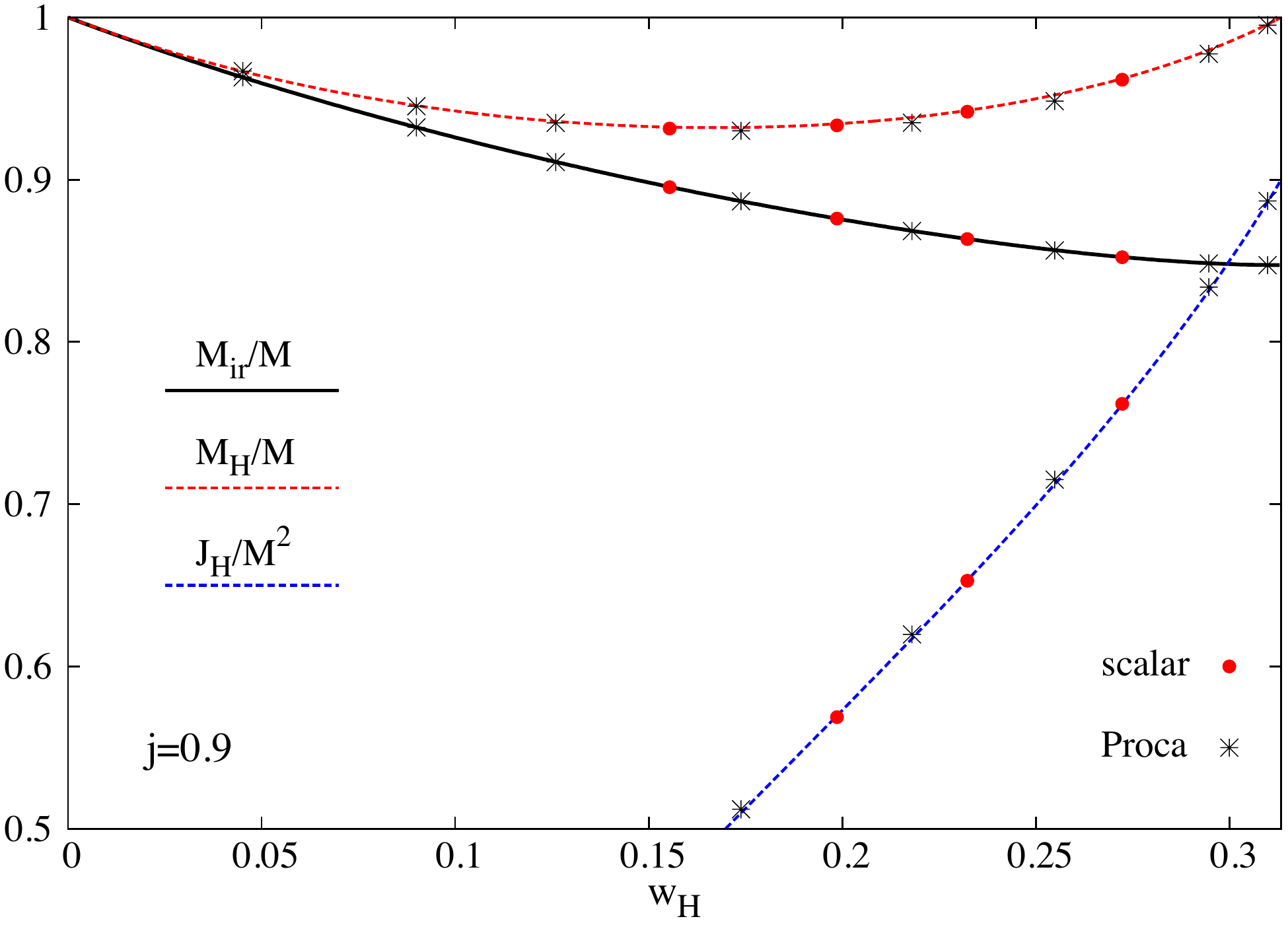}
\end{center}
\caption{\small Normalised irreducible mass, horizon mass and angular momentum: numerical solutions (points) $vs.$ the analytic model (curves). Here we have taken $j=0.9$ (in~\cite{East:2017ovw} the authors took $j=0.99$) to show the universality of the agreement.}
\label{comparisonEP}
\end{figure}

%%%%%%%%%%%%%%%%%%%%%%
%%%%%%%%%%%%%%%%%%%%%%
\noindent{\bf {\em \textit{Dynamical formation from superradiance.}}} 
%%%%%%%%%%%%%%%%%%%%%%
%%%%%%%%%%%%%%%%%%%%%%
The qKH model was suggested  by a phenomenological observation in~\cite{East:2017ovw}. In Fig. 3 therein, a Kerr-like approximation was observed to accurately fit the equilibrium state obtained from the numerical simulations of the growth of the superradiant instability, triggered by a Proca field. To establish that such equilibrium states are the hairy BHs with a \textit{quasi}-Kerr horizon that we have described before, Fig.~\ref{comparisonEP} below mimics Fig. 3 in~\cite{East:2017ovw}, wherein the normalised irreducible mass, $M_H$ and $J_H$ are plotted. In the formalism introduced above, these quantities translate as  $M_{\rm ir}/M=\sqrt{a_H}$, $M_H/M=1-p$ and $J_H/M^2=j(1-q)$. 

Fig.~\ref{comparisonEP}  shows that the qKH model fits accurately \textit{our} solutions of both Kerr BHs with scalar or Proca hair, as it did fit the equilibrium BHs obtained through numerical evolutions \textit{by East and Pretorius} in~\cite{East:2017ovw}. This establishes that the equilibrium states obtained through the growth of the superradiant instability are the fundamental states of Proca BHs with synchronised hair, $cf.$ the Appendix, and in particular the ones with a \textit{quasi-}Kerr horizon. It also \textit{predicts} a similar result for the scalar case: superradiance forms BHs with synchronised scalar hair and a \textit{quasi-}Kerr horizon (see~\cite{Brito:2014wla} for related observations).

We can now investigate how the dynamics of superradiance \textit{migrates} a vacuum Kerr BH into a BH with synchronised hair. This is exhibited in Fig.~\ref{migration}, for the examples of the numerical evolutions in~\cite{East:2017ovw}. Under the assumption used therein that a \textit{single} superradiant mode is present (the fastest growing mode), which implies that axisymmetric is a good approximation during the evolution, radiation is negligible~\cite{East:2017ovw} and the total mass and angular momentum are preserved. The process is thus~\textit{conservative}. This implies that the migration in Fig.~\ref{migration} -- an ADM mass $vs.$ horizon angular velocity plot -- occurs along a horizontal line, ending at a hairy BH with the same $j$ as the initial Kerr solution. We have verified there is a \textit{unique} such solution -- $cf.$ the inset in~Fig.~\ref{migration}. In particular this implies the equilibrium BH with synchronised hair has $j\leqslant 1$. Solutions with $j\leqslant 1$ exist in a sub-strip of the allowed strip $p<0.3$.  This explains our emphasis in this region in Fig.~\ref{colours}. Using eq.~\eqref{expr1} one shows, moreover, that $j\leqslant 1\Rightarrow p\lesssim 0.0973$, a stronger bound on the hairiness that can form dynamically from superradiance.

\begin{figure}[ht]
\begin{center}
\includegraphics[width=0.48\textwidth]{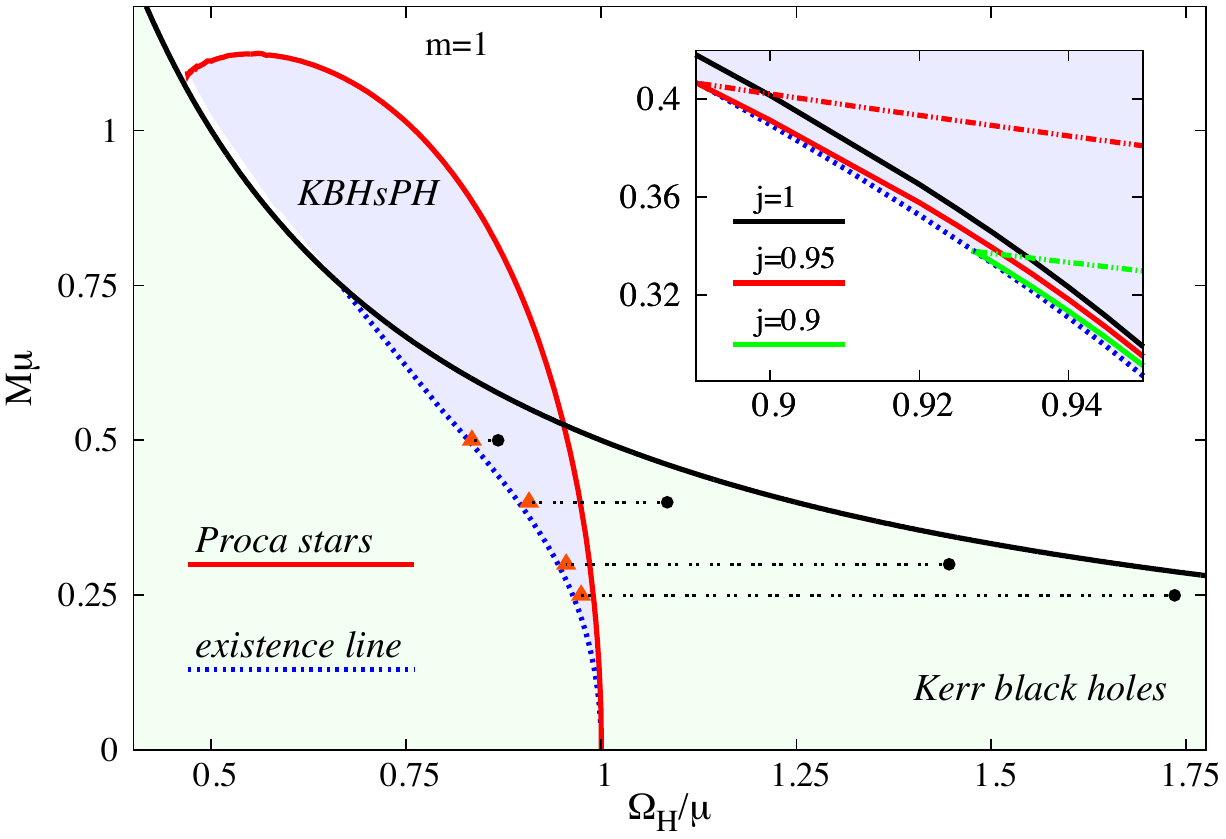}
\end{center}
\caption{\small (Main panel) Domain of existence of fundamental states of KBHsPH (shaded blue region). Vacuum Kerr BHs exist below the black solid line (corresponding to extremal Kerr). The horizontal dotted lines show the migration trajectories from vacuum Kerr BHs (black dots) to hairy BHs (red triangles) of the evolutions in~\cite{East:2017ovw}. The inset shows constant $j$ lines for both vacuum Kerr (dashed) and hairy BHs (solid), which always meet at the existence line, for any $j<1$. Migration of a Kerr BH with spin $j$ terminates when the horizontal line (constant $M$) meets a hairy BH with that  $j$ value.}
\label{migration}
\end{figure}

%%%%%%%%%%%%%%%%%%%%%%
%%%%%%%%%%%%%%%%%%%%%%
\noindent{\bf {\em \textit{Hairy BHs are entropically favoured.}}} 
%%%%%%%%%%%%%%%%%%%%%%
%%%%%%%%%%%%%%%%%%%%%%
The (vacuum Kerr) $\rightarrow $ (hairy BH) migration is conservative (in the above description), but it is irreversible. Thermodynamics determines the arrow of time. To understand this, we resort to the qKH model, which allows us to show that the hairy BH is always entropically favoured (as observed in~\cite{Herdeiro:2014goa} from the numerical data). Indeed, from the $a_H(j,p)$-cubic equation:
%\begin{equation}
%\label{aH-new}
$a_H^3-2(1-p)a_H^2+
\left[
{j^2}/{4}+(1-p)^2
\right]a_H={j^2(1-p)^2}/{4} ,$
%\end{equation} 
the solution for small $p$ [up to $\mathcal{O}(p^3)$] reads
\begin{equation}
\label{small-p}
a_H=a_H^{(Kerr)}+
\left[
\frac{(1+\sqrt{1-j^2}-\frac{1}{2}j^2)(1+\sqrt{1-j^2})}{j^2\sqrt{1-j^2}}
\right]p^2 \ .
\end{equation} 
Thus for the same $M,J$, the hairy BH is entropically favoured over the Kerr BH. In order words, the direction of the migration in Fig.~\ref{migration} is determined by the second law of thermodynamics.

%%%%%%%%%%%%%%%%%%%%%%
%%%%%%%%%%%%%%%%%%%%%%
\noindent{\bf {\em \textit{Remarks.}}} 
%%%%%%%%%%%%%%%%%%%%%%
%%%%%%%%%%%%%%%%%%%%%%
It was observed in~\cite{Herdeiro:2014goa} that there is a non-uniqueness for Kerr BHs with synchronised scalar hair (as there is in the Proca case). Near the Kerr limit, this degeneracy is discrete and of degree two: fixing  $M,J$ there exists a hairy BH and a vacuum Kerr BH. The numerical simulations reported in~\cite{East:2017ovw}, together with the qKH analytic model we have proposed and the fundamental states of KBHsPH described in the Appendix,  support the conclusion that these degenerate states correspond to the \textit{initial} (Kerr) and \textit{final} (hairy) states of the superradiant instability.

When more than one mode (and with different values of $m$)  becomes important during the superradiant evolution, the axi-symmetry assumption in~\cite{East:2017ovw} may not be accurate and less smooth evolutions may occur, $cf.$ the results reported in non-asymptotically flat setups~\cite{Sanchis-Gual:2015lje,Bosch:2016vcp,Sanchis-Gual:2016tcm}.  In this case, BHs with synchronised hair and a quasi-Kerr horizon may be transient equilibrium states. One may imagine, for instance, that the equilibrium states dynamically attained in~\cite{East:2017ovw}, with $\omega_{\rm eq}/m=\Omega_H^{\rm eq}$, may be driven to evolve by a lower frequency modes, $\omega_{\rm new}/m<\Omega_H^{\rm eq}$.   An evolution of $\Omega_H^{\rm eq}$ towards synchronising it with $\omega_{\rm new}$ is, however, hampered by the existing $\omega_{\rm eq}$ mode. Understanding this process is a relevant open issue~\footnote{This process is dissipative and likely more dissipative in the Proca case~\cite{East:2017mrj}.}.

Finally, it would be interesting to adapt the qKH model for higher dimensional BHs with synchronised hair~\cite{Brihaye:2014nba,Herdeiro:2015kha}, including in $AdS$~\cite{Dias:2011at}. Another possible application of this model is in a study of the thermodynamics of BHs with synchronised hair.

\bigskip
%\newpage

%%%%%%%%%%%%%%%%%%%%%%%%%%%%%%%%%%%%%%%%%%%%%%%%%%%%%%%%%%%%%%%%%%%%%
\noindent{\bf {\em Acknowledgements.}}
%\noindent{\bf{\em Acknowledgements.}}
%%%%%%%%%%%%%%%%%%%%%%%%%%%%%%%%%%%%%%%%%%%%%%%%%%%%%%%%%%%%%%%%%%%%%
We are very grateful to W. East and F. Pretorius for correspondence and V. Cardoso, P. Cunha and P. Pani for comments. C. H. and E. R. acknowledge funding from the FCT-IF programme.  This work was partially supported by
the H2020-MSCA-RISE-2015 Grant No. StronGrHEP-690904,  and by the CIDMA project
UID/MAT/04106/2013. Computations were performed at the Blafis cluster, in Aveiro University.

\bigskip

%%%%%%%%%%%%%%%%%%%%%%
%%%   REFERENCES   %%%
%%%%%%%%%%%%%%%%%%%%%%

\bibliography{letter_shadows}

\bigskip

%%%%%%%%%%%%%%%%%%%%%%%%%%%%%%%%%%%%%%%%%%%%%%%%%%%%%%%%%%%%%%%%%%%%%
\appendix\noindent{\bf  Appendix.} {\bf {\em{Fundamental states of spinning Proca stars and of KBHsPH.}}}
%%%%%%%%%%%%%%%%%%%%%%%%%%%%%%%%%%%%%%%%%%%%%%%%%%%%%%%%%%%%%%%%%%%%%
In this Appendix we shall briefly describe the existence of the  \textit{fundamental states} of KBHsPH that have been used in the main text. In order to understand these solutions we first describe their solitonic limit. 

Proca stars are gravitating solitons, obtained as solutions of the Einstein-Proca system (minimally coupled and without self-interactions)~\cite{Brito:2015pxa}. They are, in many ways, similar to the better known (scalar) boson stars~\cite{Schunck:2003kk,Liebling:2012fv}. The latter ones admit a discrete set of solutions, labelled by the number of nodes, $n\in \mathbb{N}_0$, of the scalar field profile function. The solutions with $n=0$ are called \textit{fundamental states}, whereas the ones with $n>0$ are called \textit{excited states}. This holds for  both spherically symmetric and axisymmetric (rotating) boson stars. In the spherical case, excited states of boson stars are known to decay to fundamental states~\cite{Balakrishna:1997ej}.

In the case of Proca stars, one expects a similar structure of fundamental and excited states, but their precise definition is less clear, due to the fact that there are (in general) four functions defining the Proca potential profile, $\mathcal{A}$, as compared to a single profile function in the scalar case. The situation is simpler in the spherical case, where the solutions are found with the ansatz~\cite{Brito:2015pxa}
\begin{equation}
ds^2=-\sigma^2(r)N(r)dt^2+\frac{dr^2}{N(r)}+r^2d\Omega_2 \ , 
\label{ansatz1}
\end{equation}
where $N(r)\equiv 1-{2m(r)}/{r}$ and 
\begin{equation}
\mathcal{A}=e^{-iwt}\left[f(r)dt+ig(r)dr\right] \ .
\label{ansatz2}
\end{equation}
which only involves two functions in defining the Proca potential. The equations of motion imply~\cite{Brito:2015pxa}
 \begin{equation}
 \frac{d}{dr}\left\{{r^2\sigma(r) N(r) g(r) } \right\}=-\frac{w r^2f(r)}{\sigma(r) N(r)} \ .
 \label{proca2}
 \end{equation}
 Integrating the left hand side of~\eqref{proca2}, between $r=0$ and $r=\infty$, and taking into account the asymptotic behaviour of the functions $\sigma(r)$, $m(r)$ and $g(r)$, $cf.$~\cite{Brito:2015pxa}, one shows the integral vanishes. Since $\sigma(r)$ and $N(r)$ have no zeros (due to the absence of horizons), this implies that $f(r)$ must have at least one zero.  Thus, for spherically symmetric Proca stars, the temporal component of the Proca 4-potential must have at least one node. We observe that, nevertheless, some of the spherical Proca star  solutions are stable \cite{Brito:2015pxa,Sanchis-Gual:2017bhw}.
 
 Rotating boson stars were found in~\cite{Brito:2015pxa} with the ansatz
 \beq
  ds^2=-&& e^{2F_0(r,\theta)}dt^2+e^{2F_1(r,\theta)}(dr^2+r^2 d\theta^2) \nn\\
 && +e^{2F_2(r,\theta)}r^2 \sin^2\theta \left(d\varphi-\frac{W(r,\theta)}{r}dt\right)^2 \ , \nonumber
\eeq
and
\begin{equation}
\mathcal{A}=\left( iVdt   +
\frac{H_1}{r}dr+H_2d\theta+i H_3 \sin \theta d\varphi 
\right)
e^{i(m\varphi-w t)} \, , \ \ \ \ \ 
\label{pra}
\end{equation}
 with $m\in \mathbb{Z}^+$. All four functions $(H_i,V)$ depend also on $r,\theta$.
 
\begin{figure}[ht]
\begin{center}
\includegraphics[width=0.48\textwidth]{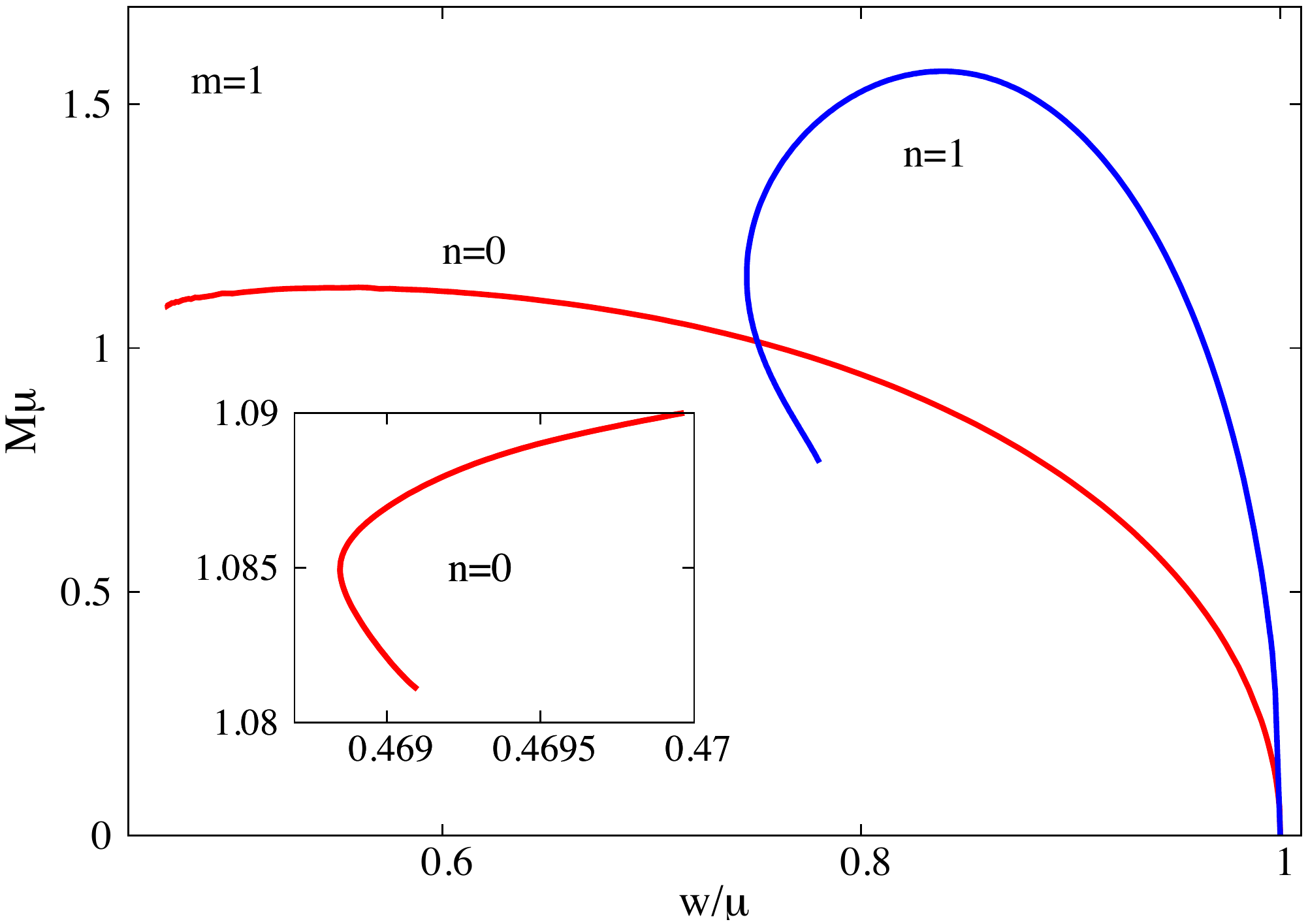}
\end{center}
\caption{\small Fundamental (red, $n=0$) and excited (blue, $n=1$) states of rotating Proca stars. Observe that excited states attain higher ADM mass, whereas fundamental states exist for a wider range of the Proca field frequency. The inset zooms in around the small backbending obtained for the $n=0$ curve, demonstrating that we have reached the end of the first branch.
}
\label{appendix1}
\end{figure}

 Regarding the definition of excited states, an analogous argument for the minimum number of nodes of $\mathcal{A}_t$ \textit{does not} exist for the axisymmetric case ($cf.$ Sec. 4 in~\cite{Herdeiro:2016tmi}). 
Still, given the mathematical proof in the spherical case that fundamental states of Proca stars have one node for $\mathcal{A}_t$,
 the rotating Proca stars reported both in~\cite{Brito:2015pxa} and in~\cite{Herdeiro:2016tmi} all had one node. 
It turns out, however, that \textit{rotating Proca stars with zero nodes of the $\mathcal{A}_t$  
exist and are the true fundamental states of Proca stars}.
 We have now been able to find these solutions, 
using similar techniques to those described in~\cite{Brito:2015pxa,Herdeiro:2016tmi},
 which indeed have lower energy. 
The ADM mass of fundamental states or rotating Proca stars ($n=0$), together with the first excited states previously reported in~\cite{Brito:2015pxa,Herdeiro:2016tmi} ($n=1$) are exhibited in Fig.~\ref{appendix1}.

\begin{widetext}

\begin{figure}[ht]
\begin{center}
\includegraphics[width=0.48\textwidth]{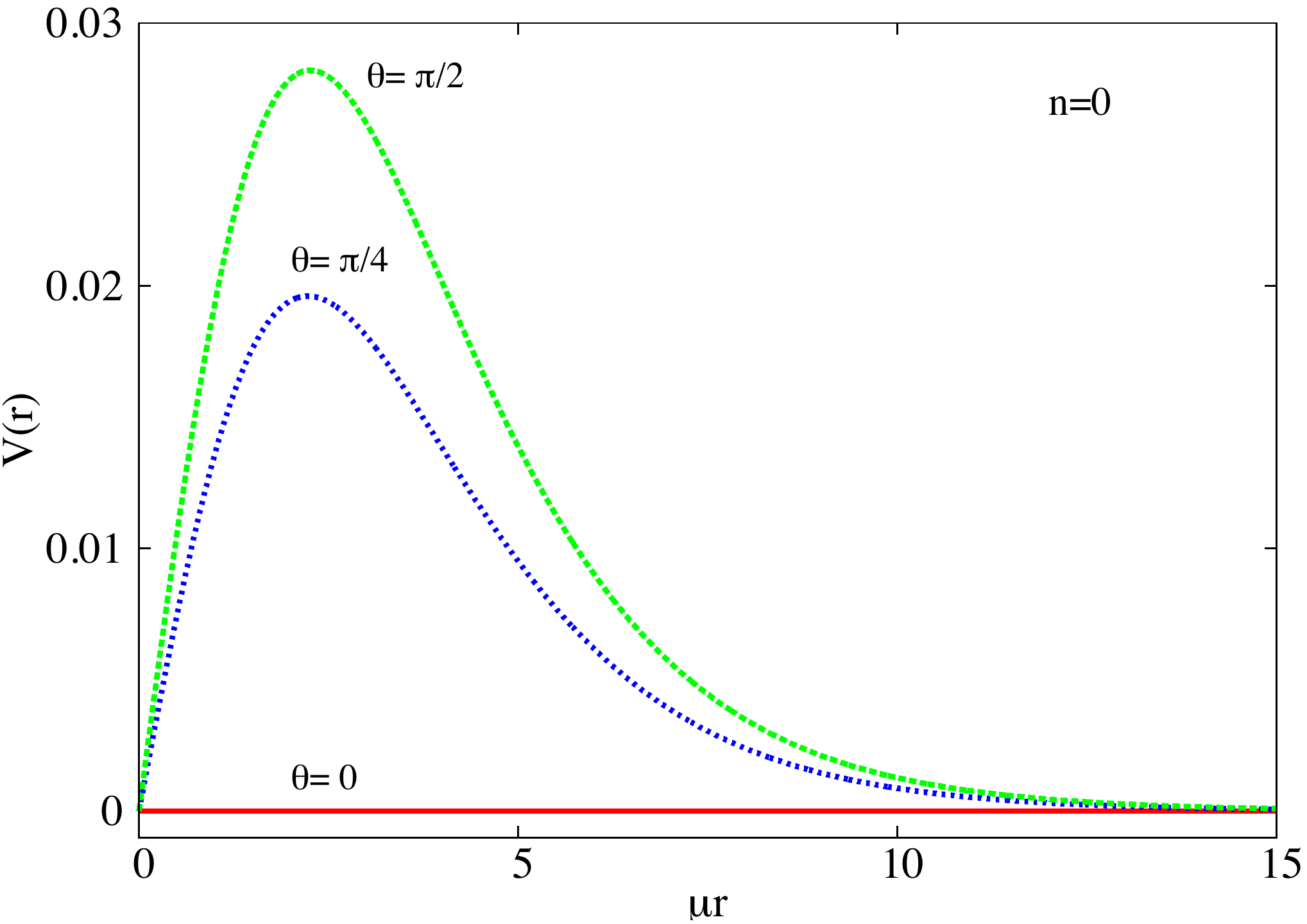}
\includegraphics[width=0.48\textwidth]{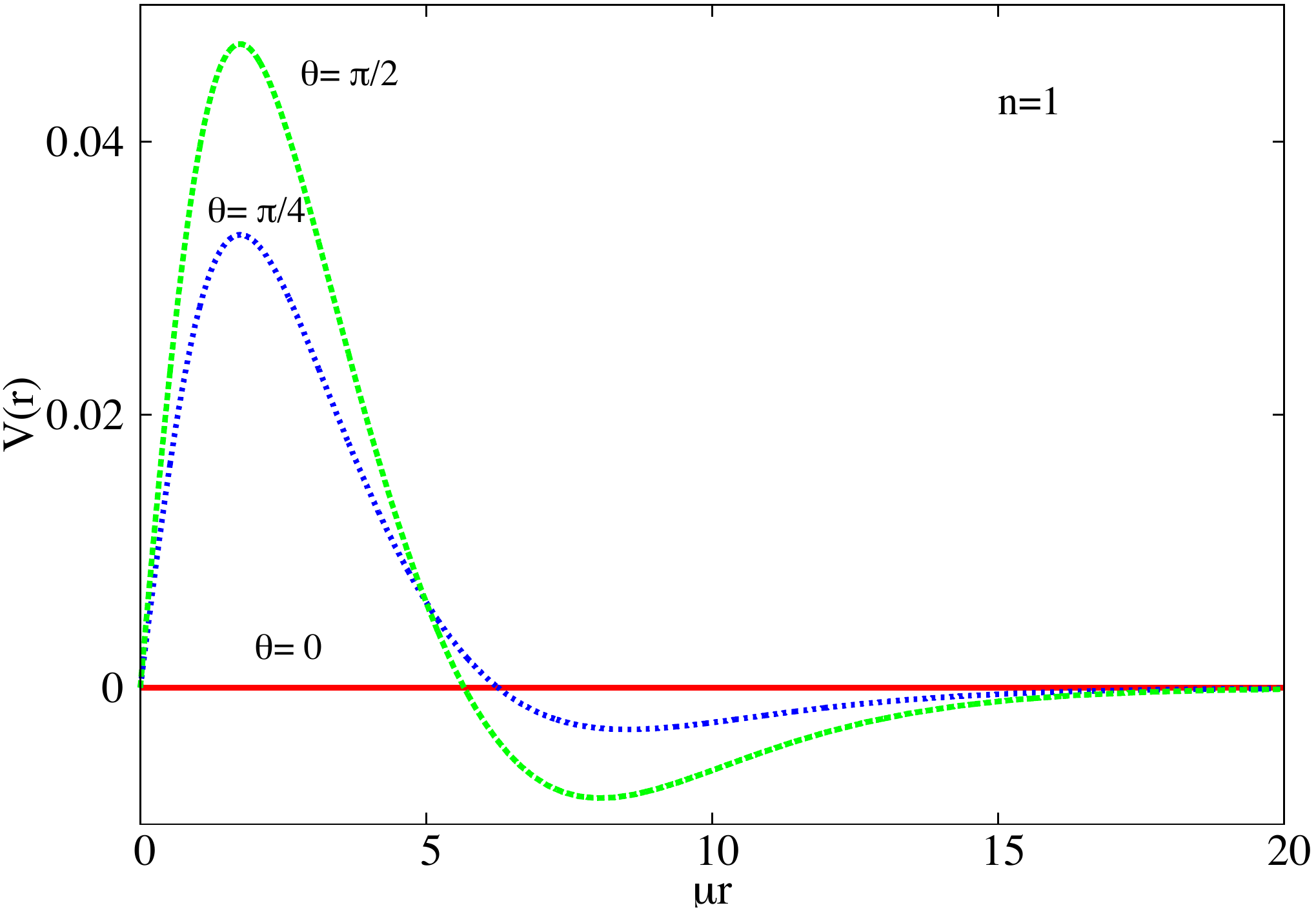}
\end{center}
\caption{\small Radial profile, for three different values of the angular coordinate $\theta$, of the function $V(r,\theta)$ that defines the temporal component of the Proca potential for rotating Proca stars. The left (right) panel is for a fundamental (excited) state with $n=0$ ($n=1$), both with $w/\mu=0.85$. Note that this function vanishes along the $z-$axis in both cases.}
\label{appendix2}
\end{figure}

\end{widetext}

To illustrate the radial profile of $\mathcal{A}_t$ we exhibit in Fig.~\ref{appendix2} the function $V(r,\theta)$, $cf.$~\eqref{pra}, for both a fundamental Proca star (left panel) 
and an excited one (right panel). The difference in the nodes structure is clearly visible. Moreover, this pattern is shared also by the magnetic potentials $H_i$, which are nodeless for $n=0$ and have one node for $n=1$.

It is now clear that in the same way that the rotating Proca stars reported in~\cite{Herdeiro:2016tmi} 
are $n=1$ excited states, so are the hairy BHs continuously connected to these Proca stars. 
The \textit{fundamental states} of KBHsPH are continuously connected to the fundamental states of ($n=0$) Proca stars, which form a part of the boundary of their domain of existence.
 Moreover, they preserve the node structure for $\mathcal{A}$ found in the solitonic limit. 
Using a similar strategy to that in~\cite{Herdeiro:2016tmi} we have spanned the domain of existence of these fundamental states of KBHsPH.  Qualitatively, this is similar to the one obtained in~\cite{Herdeiro:2016tmi} for the excited states of KBHsPH (see Fig. 6 therein) and it is exhibited in Fig.~\ref{migration}. 
Since our main focus in this paper is the region close to Kerr, wherein the qKH model applies, we shall postpone a detailed analysis of the full domain of existence, as well as specific properties of the solutions to elsewhere.  
 Finally, we note that the 
$n=1$ KBHsPH  reported in \cite{Herdeiro:2016tmi} 
satisfy the predictions of the analytic model in this work
with roughly the same accuracy as in the $n=0$ case.

%%%%%%%%%%%%%%%
%%%   END   %%%
%%%%%%%%%%%%%%%
 
\end{document}